# M3R: Increased Performance for In-Memory Hadoop Jobs

Avraham Shinnar
IBM Research
Hawthorne, NY
shinnar@us.ibm.com

David Cunningham
IBM Research
Hawthorne, NY
dcunnin@us.ibm.com

Benjamin Herta
IBM Research
Hawthorne, NY
bherta@us.ibm.com

Vijay Saraswat
IBM Research
Hawthorne, NY
vsaraswa@us.ibm.com

## ABSTRACT

Main Memory Map Reduce (M3R) is a new implementation of the Hadoop Map Reduce (HMR) API targeted at online analytics on high mean-time-to-failure clusters. It does not support resilience, and supports only those workloads which can fit into cluster memory. In return, it can run HMR jobs unchanged – including jobs produced by compilers for higher-level languages such as Pig, Jaql, and SystemML and interactive front-ends like IBM BigSheets – while providing significantly better performance than the Hadoop engine on several workloads (e.g. 45x on some input sizes for sparse matrix vector multiply). M3R also supports extensions to the HMR API which can enable Map Reduce jobs to run faster on the M3R engine, while not affecting their performance under the Hadoop engine.

## 1. INTRODUCTION

The Apache Hadoop Map Reduce (HMR) engine [6, 20] has had a transformational effect on the practice of Big Data computing.[1]

HMR is modeled on the Google Map Reduce programming model [12] (and the backing Google File System, [14]). Usually input is taken from (and output is written to) a distributed, resilient file system (such as HDFS). A partitioned input key/value (KV) sequence $I$ is operated on by *mappers* to produce another KV sequence $J$, which is then sorted and grouped ("shuffled") into a sequence of pairs of key/list of values. The list of values for each key is then operated upon by a reducer which may contribute zero or more KV pairs to the output sequence. If the involved data sets are large, they are automatically partitioned across multiple nodes and the operations are applied in parallel.

[1]In this paper we are primarily concerned with v 0.22.* Hadoop APIs. The Map Reduce APIs of interest are the "old style" APIs `mapred` and the "new style" APIs `mapreduce`.

This model of computation has many remarkable properties. First, it is simple. The HMR API specifies a few entry points for the application programmer — mappers, reducers/combiners, partitioners, together with input and output formatters. Programmers merely need to fill out these entry points with (usually small) pieces of sequential code.

Second, it is widely applicable. A very large class of parallel algorithms (on structured, semi-structured or unstructured data) can be cast in this map/shuffle/reduce style.

Third, the framework is parallelizable. If the input data sequence is large, the framework can run mappers/ shufflers/ reducers in parallel across multiple nodes thus harnessing the computing power of the cluster to deliver scalable throughput.

Fourth, the framework is scalable: it can be implemented on share-nothing clusters of several thousand commodity nodes, and can deal with data sets whose size is bounded by the available disk space on the cluster. This is because mappers/shufflers/reducers operate in streaming mode, thus supporting "out of core" computations. Data from the disk is streamed into memory (in implementation-specified block sizes), operated on, and then written out to disk.

Fifth, the framework is resilient. A job controller tracks the state of execution of the job across multiple nodes. If a node fails, the job controller has enough information to restart the computation allocated to this node on another healthy node and knit this new node into the overall computation. There is no need to restart the entire job.[2] Key to resiliency is that the programmer supplied pieces of code are assumed to be functional in nature, i.e. when applied to the same data the code produces the same result.

Because of these properties, the HMR engine is now widely used, both as a framework against which people directly write code (e.g. for Extract/ Transform/ Load tasks) and as a compiler target for higher-level languages (cf Pig [16], Jaql [7], Hive [19], SystemML [15]).

The design point for the HMR engine is *offline* (batch) long-lived, resilient computations on large commodity clusters. To support this design point, HMR makes many decisions that have a substantial effect on performance. The HMR API supports only single-job execution, with input/ output being performed against an underlying file system

[2]Within limits; of course if there are a large number of failures, the job controller may give up. The job controller itself is a single point of failure, but known techniques can be applied to make it resilient.

(HDFS). If a higher level task is to be implemented with multiple jobs, each job in this sequence must write out its state to disk and the next job must read it in from disk. This incurs I/O cost as well as (de-)serialization cost. Mappers and reducers for each job are started in new JVMs (JVMs typically have high startup cost). An out-of-core shuffle implementation is used: the output of mappers is written to local disk; a file transfer protocol is used to move these files to their target nodes and an out-of-core sorting algorithm is used to group the records by key for the reduce phase.

The demands of *interactive analytics* (e.g. interactive machine learning) lead to a different design point. Here the data to be operated upon has already been cleaned and digested and reduced to arrays of numbers that are (on the higher end) terabytes big (rather than petabytes). This data can be held in the memory for scores of nodes (one does not need thousands of nodes). Indeed, the amount of main memory available on nodes is only going to increase in coming years, making in-memory execution even more attractive. Furthermore, algorithms in this space tend to be iterative, operating on large data-structures in phases. Performance is critical – performance closer to that delivered by in-core (multi-node) HPC algorithms is desired. This performance point can be 10x-100x better than the performance delivered by Hadoop.

***M3R.*** In this paper we make a fundamental distinction between the Hadoop Map Reduce APIs (we will call them HMR APIs, or just HMR) and the Hadoop Map Reduce *implementation* (we will call it the HMR Engine). M3R implements the HMR APIs – thus it can run existing Hadoop jobs, including jobs produced by tool-chains above Hadoop, such as Pig, Jaql and SystemML. The HMR APIs supported by M3R include the `mapred` and `mapreduce` APIs, counters, user-specified sorting and grouping comparators, user-defined input/output formats and the Hadoop distributed cache. M3R is essentially agnostic to the file system, so it can run HMR jobs that use the local file system or HDFS.

However, M3R is a completely new engine focused on the following design points:

- In-memory execution: M3R stores key value sequences in a family of long-lived JVMs, sharing heap-state between jobs. Note that this limits M3R scalability to the size of memory on the cluster (not the size of disks on the cluster).

- No resilience: The engine will fail if any node goes down – it does not recover from node failure.

  This means that M3R is useful for networks with high mean time to failure. These can be commodity clusters with scores of nodes or high performance systems with much larger node counts. This also means that M3R is probably not suitable for jobs that require tens of hours to run since such long runs might interfere with operational maintenance issues.

- Performance: The engine should deliver performance close to main memory execution.

M3R is implemented in X10 [10, 17], a type-safe, object-oriented, multi-threaded, multi-node, garbage-collected programming language designed for high-productivity, high-performance computing. X10 is built on the two fundamental notions of *places* and *asynchrony*. An X10 program typically runs as multiple operating system processes (each process is called a *place* and supplies memory and worker-threads), and uses asynchrony within a place and for communication across places. Over an essentially standard modern, sequential, class-based, object-oriented substrate (with support for functions and structs, and a sophisticated constraint-based type system), X10 has four core, orthogonal constructs for concurrency and distribution: **`async`** `S` (to execute `S` asynchronously), **`finish`** `S` (to execute `S` and wait for all its asyncs to terminate), **`when`** `(c) S` (to execute `S` in one step from a state in which `c` is true), and **`at`** `(p) S` (to switch to place `p` to execute `S`). The power of X10 arises from the fact that these constructs can be nested arbitrarily (with very few restrictions), and thus lexical scoping can be used to refer to local variables across places. The X10 compiler produces C++ for execution on a native back-end and also Java for execution on multiple JVMs. The X10 runtime (written primarily in X10) ensures that the execution of `at` transparently serializes values, transmits them across places and reconstructs the lexical scope at the target place. The X10 runtime provides fast multi-place coordination mechanisms, such as barriers and teams. X10 runs on sockets, on PAMI, and on MPI (and hence on any transport on which MPI runs, including Infiniband).

The M3R engine, implemented in X10, enjoys the following key advantages over the HMR engine:

1. Each instance of M3R runs on a fixed number (possibly one) of multi-threaded JVMs. An M3R instance runs all jobs in the HMR job sequence submitted to it, potentially running multiple mappers and reducers in the same JVM (for the same job), and sharing heap-state between jobs.

2. The job tracker and the heartbeat mechanism is completely eliminated. Instead fast X10 constructs (barriers, teams) are used to coordinate mappers and reducers and signal job completion.

3. On input from the file system, M3R associates the input splits with the global (multi-place) key value sequence obtained from this input. Subsequent invocations of the input splits (e.g. by subsequent jobs in the sequence) are fulfilled by reading the key value sequence from the heap, eliminating the need to read from the file system again, and deserialize.

   Similarly, output to an output formatter is associated with the global key value sequence so that subsequent input requests can be fulfilled from the key value sequence. (See Section 3.2.1.)

4. The shuffle of key value pairs is done in memory, using X10 inter-process communication. It enjoys the benefit of de-duplication performed by the X10 serialization mechanism. (See Section 3.2.2.3.)

5. M3R implements a guarantee (*partition stability*) that the same partition number is mapped to the same place, across all jobs in the sequence. This crucial property enables programmers to write HMR jobs that can simply re-use memory structures across jobs and avoid a significant amount of communication. (See Section 3.2.2.2.)

The payoff from these advantages is that small HMR jobs can run essentially instantly on M3R, avoiding the huge (10s of second) start-up cost of the HMR engine. We show that some HMR jobs can run 50x faster on M3R than on the HMR engine. We show that programs in languages higher in the Hadoop tool stack (particularly Pig, Jaql and System ML jobs) can run unchanged (minor modifications are needed to the compilers for these languages).

To implement M3R we solved several technical problems:

- Control over cloning. The HMR API allows the reuse of keys and values, necessitating expensive cloning by the M3R engine. We allow the programmer to specify that keys/values need not be cloned.
- Control over caching. Allows the programmer to specify which files should be cached and when the cache should be flushed.
- Exposing partition stability, allowing for locality-aware programming within the confines of the HMR API.

In summary the contributions of this work are:

1. Identifying a distinction between HMR APIs and the HMR engine.
2. Showing that HMR APIs can be implemented in a main-memory implementation with substantial benefits (reduced start-up time, significant performance gain).
3. Identifying the sources of performance gain in the M3R engine.
4. Identifying extensions to the HMR APIs which can be used by M/R jobs to achieve better performance on M3R, without affecting performance on the HMR engine.
5. Demonstrating that Pig, Jaql and SystemML programs can run unchanged on M3R (with minor modifications needed to the compilers).

***Rest of this paper.*** In the next section we place this paper in the context of related work. Section 3 discussed the basic HMR engine and M3R engine execution flows, highlighting how M3R differs from HMR. Section 4 discusses backwards compatible changes to HMR APIs that permit new kinds of input and output formatters to be written that are aware of M3R caching, and that permit information to be specified that does not affect the HMR engine but that can be used by M3R for better performance. Section 6 compares the performance of the HMR and M3R engines on several benchmark problems, and also discuss micro-benchmarks that highlight the differences between the two. Finally, Section 7 concludes and points directions for future work.

## 2. RELATED WORK

HaLoop[9] adds to the Hadoop API, allowing a programmer to explicitly specify which MapReduce iterations make up a job, and the data that is re-used across iterations. The HaLoop scheduler uses this knowledge to cache the reused data on local disks. By leveraging this data locality in the scheduler, performance is improved. The M3R engine does not extend the Hadoop API in this way, but is still able to take advantage of data locality across MapReduce iterations. The M3R engine caches all key/value pairs in memory at each machine, and it attempts to execute MapReduce jobs where their data is located. Because M3R caches everything, there is no need for an API to specify what is to be reused. Thus M3R is able to achieve the iterative job performance improvements of HaLoop, without the burden of changing existing Hadoop programs to use new APIs.

Twister[13] is another MapReduce engine that exploits data-locality for iterative MapReduce jobs. While Twister and M3R make similar design decisions, the primary difference is that M3R implements the HMR API, while Twister does not. Another less significant difference is that M3R leverages X10's built-in communications while Twister relies on an externally hosted Message Broker network.

The Spark[21] engine primarily uses main memory storage to achieve faster performance compared to Hadoop. M3R also uses main memory to achieve performance gains, but at the expense of longer recovery time in the event of a node failure. Spark uses a programming model based on Resilient Distributed Datasets (RDDs), to reduce this recovery time without the performance degradation that comes with checkpointing. While very interesting, Spark is not an HMR based engine.

The Hadoop Online Prototype(HOP)[11] engine has many similarities to M3R. Both are capable of running unmodified Hadoop jobs, and both engines speed up execution by getting data from the map phase to the reduce phase (and potentially on to the map phase of the next job) more quickly. HOP accomplishes this by pipelining the mappers and reducers, moving data between them using main-memory buffers and network links in place of the local filesystem whenever possible. M3R also uses main-memory and network links for this shuffle, but M3R goes a step further, by merging map and reduce roles into the same process. This enables M3R to reduce the amount of data moved in the shuffle phase (and in the map phase of the following job), because much of the input for the reduce phase is already in memory from the just-completed map phase of each process.

In the commercial space, companies such as Platform[5], Facebook[8], MapR[3], IBM[1], and others have focused primarily on improving Hadoop's ability to scale, eliminating single points of failure like the NameNode, and improving the manageability and support of large Hadoop clusters. Some of these implementations offer minor performance improvements, but their main focus is on the largest-scale Hadoop users. M3R's focus is on the smaller scale, on the user who finds themselves scaling down their Hadoop application size to reach completion times suitable to an interactive user. M3R offers interactive performance levels for larger data sizes than Hadoop.

## 3. HMR AND M3R EXECUTION FLOWS

This section presents a detailed account of HMR and M3R execution flows, highlighting their different approach to MR execution.

Figure 1 depicts an example application for iteratively multiplying together a sparse matrix and dense vector. (This is the core computation inside PageRank.) A single matrix multiply is implemented with two MR jobs: the first to calculate the appropriate scalar products and the second to sum them. Since the HMR API does not represent work-flows the client must submit two MR jobs (for each iteration), using the output of the first as an input to the second. Note that the input matrix `G` remains unchanged through both jobs.

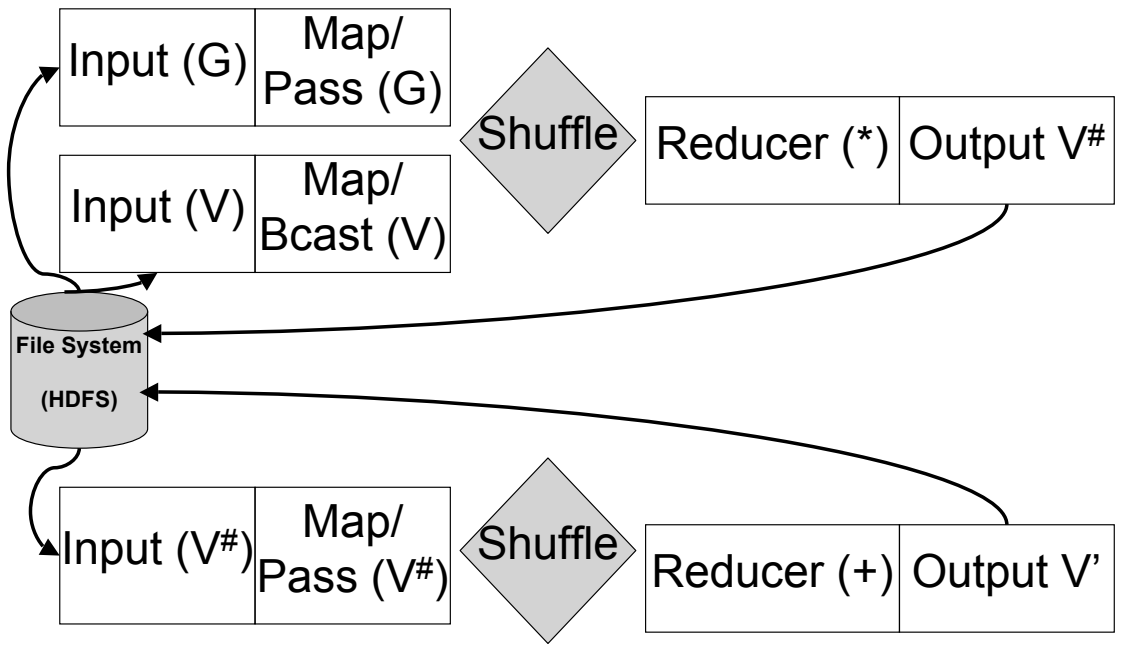

**Figure 1: Data flow for an iteration of matrix-vector multiplication. The matrix $G$ is row-block partitioned, and the vector $V$ is broadcast as needed.**

## 3.1 HMR Execution Flow

We discuss in detail the execution flow for a single job as background for a discussion of the differences with M3R.

The client prepares a job configuration object specifying the classes to be used during execution, the number of reducers used to run the job, the location of the HMR jobtracker etc. This configuration object is threaded throughout the program (and passed to user classes), and can hence be used to communicate information of use to the program.

The job configuration object is submitted in a call to `JobClient.submitJob`. This library function obtains a jobid from the Hadoop jobtracker, and writes out the necessary job information to the jobtracker's filesystem (in a jobid-relative path), including the job configuration object and the user code to be run. The user's `InputFormat` is instantiated, and asked to produce `InputSplits`, metadata that describes where each "chunk" of input resides. These are also written out to the job's directory. Finally, the jobtracker is notified that a new job with the given jobid has been submitted.

Figure 2 presents a high level view of the data flow for a single Hadoop job (each mapper and reducer box represents multiple processes).

The jobtracker schedules the job to run, allocating map and reduce tasks on available task trackers. The map tasks (allocated close to their corresponding `InputSplits`) must next read input data. If the data is in HDFS (common case), reading requires network communication with the namenode (storing the file metadata). Reading the actual data requires file system I/O (which may not require disk i/o if the data is in kernel file system buffers), and may require network i/o (if the mapper is not on the same machine as the one hosting the data). The map tasks deserialize the input data to generate a stream of key/value pairs that is passed into the mapper. The mapper outputs key/value pairs, which are immediately serialized and placed in a buffer. While in the buffer, Hadoop may run the user's combiner to combine values associated with equivalent keys. When the buffer fills up, they are sorted and flushed out to local disk.

Once map output has been flushed out to disk, reducer tasks start fetching their input data. This requires disk and network I/O. Each reducer performs an out-of-core sort of its input data. After all of the mappers have completed and the data is sorted, each reducer starts processing its input. Each reducer outputs a (possibly empty) sequence of key/value pairs that is sent to the `OutputFormat` (and its attendent `RecordWriter`) for output. Typically, this involves serializing the data and writing it out to disk. The namenode is contacted to update the required filesystem metadata. The data is written out to the local datanode (generally co-located with the compute node), and optionally replicated to a configurable number of other datanodes for resilience.

## 3.2 M3R Execution Flow

The general flow of M3R is similar to the flow of the HMR engine. The client submits multiple jobs to the M3R engine, which distributes the work to compute nodes in the cluster.

An M3R instance is associated with a fixed set of JVMs (spawned by the X10 runtime, one per place) that are used to run both mapper and reducer jobs, and is used to run multiple jobs. Reusing VMs reduces startup cost and permits data to be kept in memory between jobs. In the common case of job pipelines (the output of one job is immediately used by the next job) M3R affords significant benefits in avoiding network, file i/o and (de-)serialization costs.

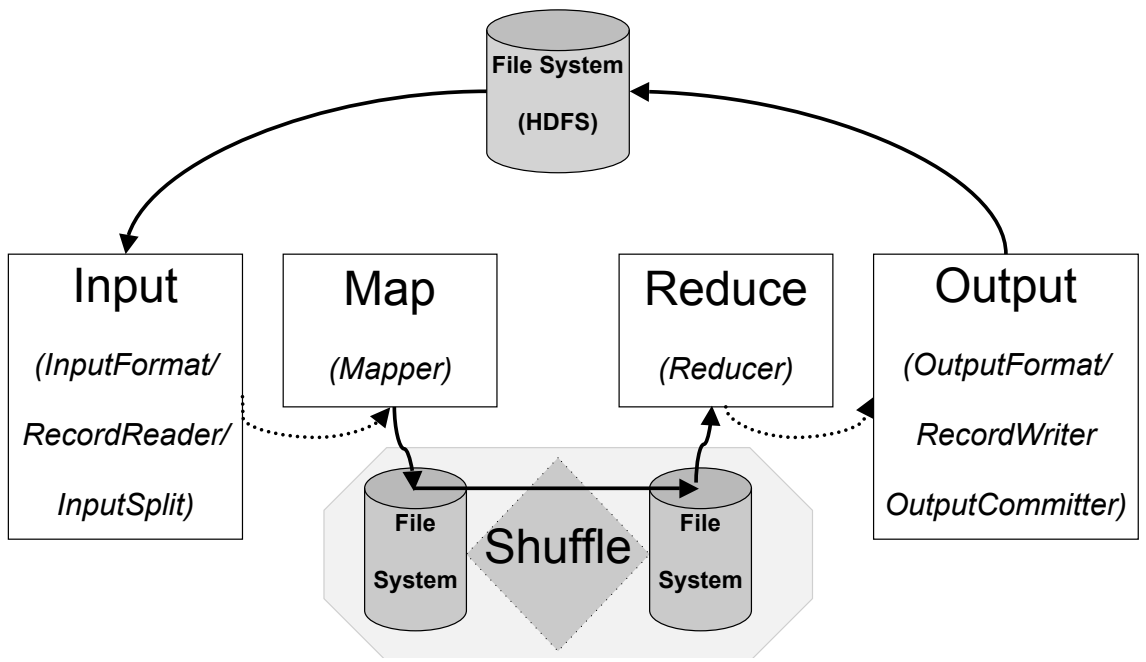

**Figure 2: Data flow in a Hadoop job. Dotted lines represent cheap in-memory communication. Solid black lines represent expensive out of memory (disk or network) operations.**

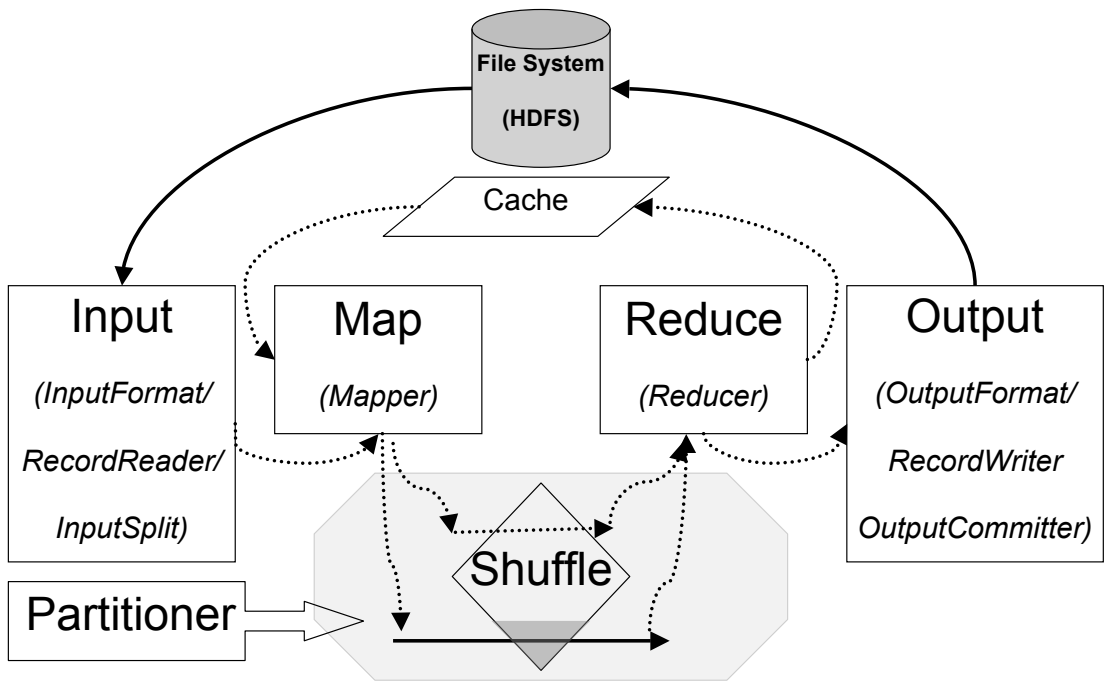

**Figure 3: Data flow in an M3R job. Dotted lines represent cheap in-memory communication. Solid black lines represent expensive out of memory (disk or network) operations.**

### 3.2.1 Input/Output Cache

M3R introduces an in-memory key/value cache to permit in-memory communication between multiple jobs in a job sequence. Like HMR, M3R can uses the client provided `RecordReader` to read in data and deserialize it into a key/value sequence. Before passing it to the mapper, M3R caches the key/value pairs in memory (associated with the input file name). In a subsequent job, when the same input is requested, M3R will bypass the provided `RecordReader` and obtain the required key/value sequence directly from the cache. As the data is stored in memory, there are no attendent (de)serialization costs or disk/network I/O activity.

Similarly, when the reducers emit a key/value pair, M3R caches it (associated with the output file name) before using the `RecordWriter` to serialize it and write it to disk. Subsequent reads from this file name can be fulfilled from the cache. If the output data is determined to be temporary (only needed for subsequent jobs in the sequence), then the data does not even need to be flushed to disk.

The cache in M3R is mostly transparent to the user, as it is intended to work with unmodified Hadoop jobs. To this end, M3R intercepts calls to the base Hadoop filesystem and attempts to keep the cache up to date. For example, deleting a file from the filesystem causes it to transparently removed from the cache.

There are some instances where cache awareness is beneficial/required. Explicit interaction with the cache is possible through a set of extensions described in Section 4.2.

### 3.2.2 Reducing Shuffle Overhead

#### 3.2.2.1 Co-location.

When M3R runs a job, it distributes the required information to each place and starts multiple mappers and reducers in each place (running in parallel). As a result, some of the data a mapper is sending is destined for a reducer running in the same JVM. The M3R engine is careful to make this case efficient, and guarantees that no network, or disk I/O is involved.

For locally shuffled data the engine tries to avoid the time and space overhead of (de)serialization. Unfortunately, due to a limitation in the Hadoop API, this cannot in general be done safely. Hadoop assumes that map (and reduce) output is immediately serialized and encourages clients to mutate them after they have been passed to the engine. This forces M3R to conservatively make a copy of every key/value pair. Section 4.1 introduces an extension allowing job writers to inform the engine that a given map or reduce class is well-behaved and does not mutate values that it has output to the engine. For these classes, the key/value pair output by the mapper can be passed directly into the reducer, avoiding (de-)serialization.

#### 3.2.2.2 Partition Stability.

The co-location strategy just described yields a small performance boost to most jobs. Some fraction of data will be shuffled locally, reducing intra-job communication. However, M3R goes further, and allows carefully written algorithms to deliberately exploit locality, dramatically decreasing communication and (de)serialization costs.

The HMR API allows the programmer to control how keys are partitioned amongst the reducers. This is done via a `Partitioner` that maps keys to partitions. This is primarily done for load balancing purposes and for global sorting (across reducers). The default implementation uses a hash function to map keys to partitions. Hadoop does not, however, allow the programmer any control over where the reducer associated with a partition is run. This is deliberate, and allows Hadoop to transparently restart failed reducers on different hosts.

M3R, in contrast, provides programs with the following *partition stability* guarantee: for a given number of reducers, the mapping from partitions to places is deterministic. This allows job sequences that use a consistent partitioner to route data locally. The output of a given reducer is cached at the place where it is written. If it is read by a subsequent job, the mapper associated with that data will be assigned to the same place. If the mapper then outputs keys that map to the same partition, it is guaranteed that the key will be locally shuffled.

For illustration consider the matrix vector multiplication example. The first job has two inputs: the matrix $G$ and the vector $V$. The matrix $G$ is far larger, as its size is quadratic in the size of $V$. As a result, it is critical that $G$ not be moved. Parts of $G$ (e.g. a set of contiguous rows) should be read in by each place and then left there for the entire job sequence. This can be accomplished by using an appropriate partitioner (e.g. one that assigns to place $i$ the $i$th contiguous chunk of rows). This ensures that for the first job all the partial products (the product of a fragment of a row with a fragment of a column, for a given row) are co-located. The same partitioner is used for the second job and ensures that they are sent to a co-located reducer for summation. As a result, the shuffle phase of the second job in each iteration can be done without any communication.

Section 6.1 discusses the (substantial) performance impact of local vs remote shuffle through a micro-benchmark.

#### 3.2.2.3 De-Duplication.

Consider the matrix vector multiply example again. The first job must broadcast $V$ to all the reducers. However, each place has a number of reducers, say $k$.

Clearly it would be beneficial for M3R to not send $k$ copies of $V$ to each place. Note the HMR engine does not co-locate reducers, so this optimization does not make sense. M3R takes advantage of a feature of the underlying X10 serialization protocol to transparently de-duplicate the data sent to a place. If the mappers at place $P$ output the identical key or value multiple times for a reducer located at place $Q$, only one copy of the key or value is serialized. On deserialization $Q$ will have multiple aliases of that copy.

## 4. HMR API EXTENSIONS

As discussed in Section 3, M3R extends the HMR APIs in a backward compatible way for three reasons: to eliminate needless (de)serialization, to interact with the cache, and to enable locality aware algorithms.

### 4.1 ImmutableOutput

The Hadoop API assumes that mapper and reducer output is immediately serialized. As a result, it allows both the mapper and reducer code to reuse keys and values after they have been output. This reuse is intended as an optimization. Instead of allocating a new object, client code can mutate and reuse a previously output object.

```java
class Map ... {
    IntWritable one = new IntWritable(1);
    Text word = new Text();

    void map(LongWritable key, Text value,
             OutputCollector output, Reporter r) {
        String line = value.toString();
        StringTokenizer tokenizer
            = new StringTokenizer(line);
        while (tokenizer.hasMoreTokens()) {
            word.set(tokenizer.nextToken());
            output.collect(word, one);
        }
    }
}
```

```java
class Map ... implements ImmutableOutput {
    IntWritable one = new IntWritable(1);

    void map(LongWritable key, Text value,
             OutputCollector output, Reporter r) {
        String line = value.toString();
        StringTokenizer tokenizer
            = new StringTokenizer(line);
        while (tokenizer.hasMoreTokens()) {
            Text word =
                new Text(tokenizer.nextToken());
            output.collect(word, one);
        }
    }
}
```

**Figure 4: Hadoop WordCount example: original (left),** `ImmutableOutput` **(right)**

For example, Figure 4 (left) presents the mapper from a typical word count example [4].

Clearly, such reuse of output objects is not compatible with caching. To maintain integrity, M3R will by default clone keys and values produced by mappers and reducers. If the map or reduce class promises to not mutate keys and values that it has emitted – it does this by implementing `ImmutableOutput` – M3R will not clone. Note that Hadoop will simply ignore this interface, allowing the same code to be run on M3R and Hadoop. The right-hand code presents the same word count mapper example modified to satisfy the `ImmutableOutput` constraints. The modified example allocates a fresh `Text` object each time instead of reusing a single object.

Section 6.3 presents performance results demonstrating the improvement this change effects for WordCount.

With the old style "mapred" interface, the user can also provide a custom `MapRunnable` implementation to manually connect the input to the mapper. Any such custom `MapRunnable` implementation must also be marked as producing immutable output for M3R to avoid cloning.

The default `MapRunnable` implementation used by Hadoop reuses the same key/value for each input and so does not conform to the required contract for `ImmutableOutput`. This means that if the mapper just passes along the input (e.g. the identity mapper), the output will be mutated by the default implementation. M3R specially detects the default implementation and automatically replaces it with a customized version that allocates a new key/value for each input and is (appropriately) marked as `ImmutableOutput`.

## 4.2 Key/Value Cache

M3R introduces a key/value cache for job inputs and outputs. Simple Hadoop programs can transparently benefit from the cache. However, more sophisticated programs can benefit from some cache interactions. These range from interfaces that allow code to teach M3R how to better interact with their custom input/output code to code that explicitly modifies or queries the cache.

### *4.2.1 Naming Data*

The HMR `InputSplit`, `InputFormat`, and `OutputFormat` classes do not declare what name is associated with a given piece of data. Focusing on `InputSplits`, this makes it difficult to identify what data input is referring to. Without a name, there is no way to cache the data for subsequent use.

M3R understands how standard Hadoop input and output formats work, in particular the `File(Input/Output)Format` classes and the `FileSplit` class. Given a `FileSplit`, it can obtain the file name and offset information and use that to enter/retrieve the data in the cache.

For user-defined splits, M3R provides the `NamedSplit` interface, allowing the `InputSplit` to provide the necessary information. The interface defines a single method, `getName`, which returns the name to use for the data associated with the split. Alternatively, if the split is a wrapper around another split (such as described in Section 4.2.2) then it can implement the `DelegatingSplit` interface and tell M3R how to get the underlying information.

If a split does not implement one of these interfaces and is not a standard type known to M3R then M3R is forced to bypass the cache for the data associated with the split. Note that (as with the `ImmutableOutput` interface) Hadoop simply ignores these interfaces, allowing the same code to run on M3R and Hadoop.

### *4.2.2 Multiple Inputs/Outputs*

The Hadoop model only allows a single input format. Similarly, each reducer writes to a single output. For many applications, this is too restrictive. For example, the iterated matrix vector multiplication job sequence discussed in Section 3 needs two inputs: the matrix and the vector. Furthermore, these inputs are routed to two different mappers. To address this type of situation, the Hadoop libraries come with the `MultipleInputs` and `MultipleOutputs` classes to multiplex input and output.

The `MultipleInputs` class uses `TaggedInputSplit` to tag input splits so they can be routed to the appropriate base input format and mapper. The `DelegatingInputFormat` class handles instantiating the underlying record readers. As a result, it needs to be cache aware. In particular, it needs to wrap the input formats it creates with the provided `CachingInputFormat` wrapper, which adds cache awareness to a base input format.

Similarly, the `MultipleOutputs` class creates additional named record writers, allowing the reducer to output to multiple explicitly named files. As with `MultipleInputs`, this code needs to be modified to enable caching.

The necessary changes to the standard libraries are transparently done by M3R. However, if client code implements their own variant of these classes, they need to make similar modifications.

#### *4.2.3 Cache Management*

Programs can explicitly manage the cache in different ways. They can mark outputs as "temporary", such that they need not be output to disk at all. This is suitable for outputs that will be consumed by subsequent jobs and are not needed by non-map/reduce code. At the moment, this is done based on a simple naming convention: if the last part of the output path starts with a given string (which defaults to "temp") then it is treated as temporary and not written out. This string can be customized by setting a property in the job's configuration. Adding settings to the job configuration like this is common practice in Hadoop for communicating additional information to jobs. In a similar way, a list of files that should be considered temporary could be passed enumerated in a job configuration setting.

Programs can also rename and delete data from the cache. M3R alters Hadoop's `FileSystem` class so that it transparently sends calls to operations such as `rename`, `delete`, and `getFileStatus` to both the cache and the underlying file system. However, there are times when the program wants to explicitly delete (or rename) data just from the cache, without affecting the underlying file system. To support this, the `FileSystem` objects created by M3R implement and additional `CacheFS` interface. This interface provides a `getRawCache` method that returns a new `FileSystem` object. Operations on this synthetic file system object are only sent to the cache of the original `FileSystem`. So calling `delete` on the synthetic file system will delete the file from the cache without affecting the underlying file system.

#### *4.2.4 Cache Queries*

Programs can also explicitly query the cache and obtain the key/value sequence associated with a path. As just discussed in Section 4.2.3, a program can use `getRawCache` in conjunction with `getFileStatus` to check if data is in the cache and obtain its associated metadata. The `CacheFS` interface provides a `getCacheRecordReader` method that allows the program to obtain an iterator over the key/value sequence associated with a given path.

### 4.3 Partition Stability

As introduced in Section 3.2.2.2, M3R allows algorithms to exploit locality to reduce shuffle costs. The interface to this ability is primarily implicit, provided by an enhanced performance model.

M3R provides a `PlacedSplit` interface that allows an input split to inform M3R what partition the data should be associated with. Splits that implement this interface are sent to a mapper running at the place associated with that partition. This is beneficial in ensuring that the data goes to the right place at the very beginning. (Using partition stability, the programmer can ensure that it stays there for the duration of the job sequence.)

The iterated matrix vector multiplication example critically takes advantage of locality. This allows M3R to run the example far more efficiently than the HMR engine. Detailed results are presented in Section 6.2.

```
Writer createWriter(File path, BlockInfo info)
Reader createReader(File path, BlockInfo info)
void delete(File path)
void rename(File src, File dest)
PathInfo getInfo(File path)
void mkdirs(File path)
```

**Figure 5: Key/Value store API. All operations are atomic (serializable).**

## 5. IMPLEMENTATION

### 5.1 The X10 Language and Core M3R Engine

X10 is a modern OO language intended for programming multi-core (providing fine-grained concurrency), heterogeneous, and distribution (scaling to thousands of nodes). It can be compiled to C++ or Java, in which case Java classes (such as those in the Hadoop codebase) are exposed alongside X10 classes. This allows us to easily use X10's sophisticated concurrency and distribution features while integrating cleanly with existing Java code.

The core M3R engine implements a minimal map/reduce API focussing on concurrency and communication, and leaving input/output to the client. It is written in pure X10, and utilizes X10's fine-grained concurrency concurrency constructs to multi-thread the mappers / reducers, and parallelize shuffling. The actual communication is handled using X10's `at (p) S` construct, which executes `S` at place `p`, automatically serializing and transmitting variables (and heap graphs referenced there-from) captured in the enclosing scope. This serialization protocol must handle cycles in the heap, so has to recognize when a given object has been serialized before. This mechanism gives us free de-duplication, as described in Section 3.2.2.3. X10's `Team` API provides a barrier construct that the engine uses for synchronization. No reducer is allowed to run until globally all shuffle messages have been sent. By utilizing X10 at the core of M3R, we leverage a highly-performing and well-tested execution engine, and can focus on the problem of map/reduce.

### 5.2 Key/Value Store

M3R caches both job inputs and outputs as discussed in Section 3.2.1. Underneath this is a distributed in-memory key/value store that implements a file system like API. The key/value store distributes the (hierarchal) metadata across the different places used by M3R.

Figure 5 presents the basic API exposed by the key value store. All operations are atomic with respect to each other, making it simple for callers to reason about their behavior. Paths are represented by Java `File`'s, which represent abstract filesystem paths.

Like HDFS, paths can map to multiple blocks, each of which can be stored at a different place. Blocks are identified by their metadata. The key value store is generic in the type of metadata, but requires that it implement a reasonable `equals` method.

The key/value store is fully distributed: both the metadata and data are distributed across the places. Metadata is distributed using a static partitioning scheme: a path is hashed to determine where the metadata associated with that path is located. Data blocks can live anywhere: there location is specified by their metadata. The `createWriter` call will create a block at the place where it is invoked.

Each place has a handle to its own concurrent hash tables (one for the metadata and one for the data). These map full paths to their associated metadata/data. When an operation needs to modify or access an entry associated with a path, it first atomically swaps out the entry with a special lock entry (or inserts it if there was nothing there beforehand). If the entry is already a lock entry, it (carefully) swaps in a heavier weight monitor entry that it then blocks on. When the task that previously locked the entry releases the lock it will detect this and wake up the blocked task.

To ensure that operations are serializable, the implementation follows the two phase locking protocol (2PL) when acquiring locks during a task. To ensure that operations do not induce a deadlock, they follow a least common ancestor-based locking protocol. Any task that acquires a lock $l$ while holdings locks $L$ must be holding the least common ancestor of $l$ with all the locks in $L$. This suffices to ensure that deadlock cannot occur.

## 5.3 Hadoop Interop

The Hadoop-interop layer of M3R wraps a JobConf and produces an X10 job that can be run by the core M3R engine. The main job class distributes task specific data to each place when it is created. It then wraps the required Hadoop API-based user code for the engine and wraps the engine's context objects to present to the Hadoop API-based user code.

The compatibility layer is complicated by the need to support two sets of Hadoop APIs: the older `mapred` and the newer `mapreduce` interfaces. Since many classes (such as `Map`) do not share a common type, separate wrapper code must be written for both of them. The implementation supports any combination of old (mapred) and new (mapreduce) style mapper, combiner, and reducer. It also support "map-only" jobs, which are Hadoop jobs with zero reducers. Output from the mapper is sent directly to output as per Hadoop.

M3R also supports many auxiliary features of Hadoop, including counters and the distributed cache. In addition to correctly propagating user counters, M3R keeps many Hadoop system counters properly updated. M3R also supports many Hadoop administrative interfaces including job queues, job end notification urls, and asynchronous progress and counter updates.

There are currently two ways to run M3R: integrated mode and server mode. Integrated mode starts the Hadoop client under the control of M3R. M3R starts and initializes the X10 runtime across all of the designated machines and (using Java classpath trickery) replaces Hadoop's `JobClient` with a custom M3R implementation that submits jobs directly to the M3R engine. It then uses reflection to call the specified client main function. When the client submits jobs they are transparently redirected to the engine. If an (M3R-aware) client explicitly wishes to use Hadoop for a specific job, they can set a property in the submitted job configuration and the `JobClient` submission logic will invoke a Hadoop server as usual. All of the benchmarks presented in this paper were run in integrated mode.

M3R also supports a (still somewhat experimental) server mode. In this mode, M3R starts up and registers an IPC server that implements the Hadoop `JobTracker` protocol. Clients can submit jobs as usual, and the M3R server (which functions just like the normal Hadoop server) will run the job. It is possible to simply replace the Hadoop server daemon with the M3R one. It is also possible to start the Hadoop and M3R servers with different configuration files, that specify different ports. They can then coexist, and a client can dynamically choose which server to submit a job to by altering the appropriate port setting in their job configuration.

Using server mode, we have successfully run BigSheets [2], a large Hadoop based system that generates assorted jobs (many of them Pig jobs). The BigSheets system was unmodified, except that we stopped the running Hadoop server and started the M3R server on the same port.

# 6. EVALUATION

To evaluate the performance of our implementation, we have measured the total running time of several benchmark programs. Each was written to the Hadoop API, with our modest M3R-specific additions discussed in Section 3. We ran these Hadoop programs in both the standard Hadoop engine and in our M3R engine, on the same input from HDFS, and verified that they produced equivalent output in HDFS (up to floating point rounding error). The hardware used was a 20 node cluster of IBM LS-22 blades connected by Gigabit Ethernet. Each node has 2 quad-core AMD 2.3Ghz Opteron processors, 16 GB of memory, and is running Red Hat Enterprise Linux 6.2. The JVM used is IBM J9 1.6.0. When running M3R on this cluster, we used one process per host, using 8 worker threads to exploit the 8 cores.

## 6.1 Microbenchmark

To illustrate these aspects of our performance model, we wrote an Hadoop application that is parameterized to simulate an arbitrary ratio of remote / local shuffling. Randomly, weighted by this local/remote ratio, pairs are either kept local or sent to an adjacent machine (thus requiring serialization and network overhead). The benchmark has three iterations, with the output of one job being the input for the next. The results are in Figure 6.

The input to this job is 1 million pairs, each with an ascending integer for key and an array of 10000 bytes for value. The mapper, which implements ImmutableOutput, randomly decides to emit the pair with either its key unchanged or replaced with a key (created during the mapper's setup phase) that partitions to a remote host. The partitioner simply mods the integer key, and the reducer is the identity reducer.

In M3R, the output of all jobs except the final iteration are marked as temporary (not written to HDFS). The initial read and the final output of course must be written to HDFS. We explicitly delete the previous iteration's input, as it will not be accessed again and its presence in the cache wastes memory.

When running in Hadoop, every iteration takes the same amount of time, regardless of whether pairs are shuffled remotely or locally. This is because Hadoop does not provide a notion of remote / local shuffle because there is no partition stability. All shuffled data is serialized and communicated via local files and network and therefore there is equal cost for all destinations. Also, since Hadoop does not cache data between jobs, disk I/O occurs at the beginning and end of every job. The second and subsequent iterations cannot benefit from I/O performed by the first iteration, as the loaded data is not kept in memory between jobs.

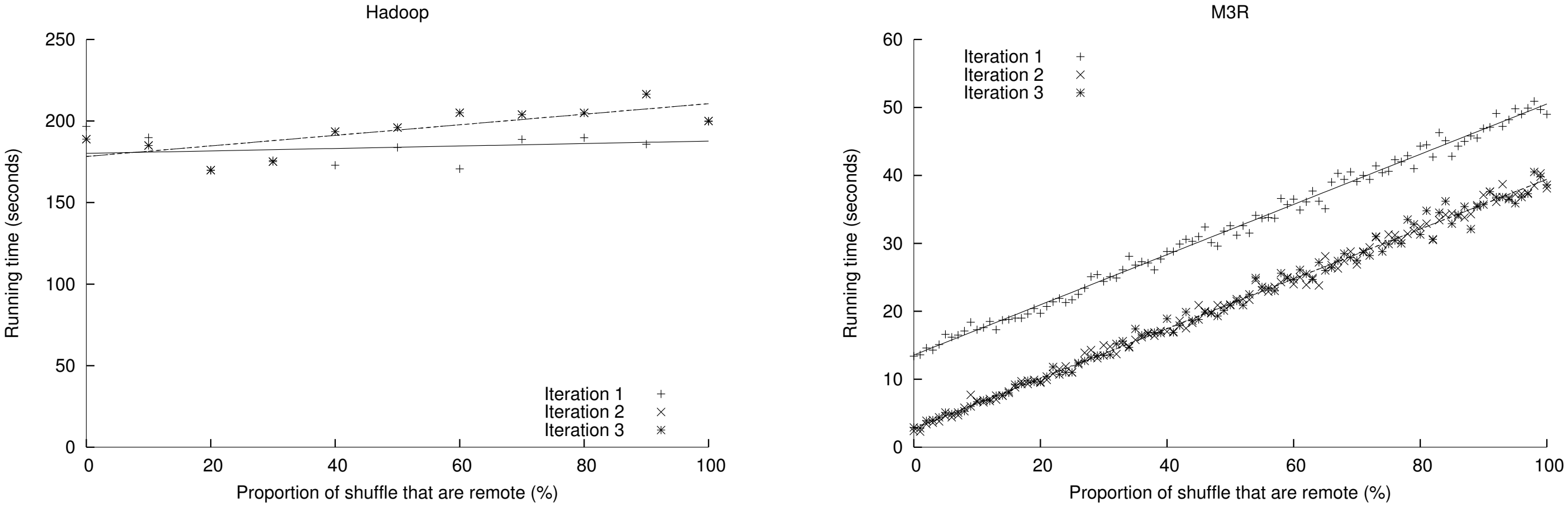

**Figure 6: Performance profile of microbenchmark in Hadoop and M3R**

When running in M3R, the performance changes drastically according to the amount of remote shuffling and due to cache hits in second and subsequent iterations. All iterations exhibit a linear relationship between the amount of remote shuffling required and the time taken, while also having a constant overhead. However the constant overhead is considerably less in the second and third iterations since pairs are fetched directly from the cache instead of being deserialized from HDFS. Note also that in M3R, even the first iteration with 100% remote shuffles outperforms Hadoop by a considerable margin. We assume this is due to overheads inherent in Hadoop's task polling model, disk-based out-of-core shuffling, and JVM startup/tear down costs.

### *6.1.1 Repartitioning*

Because we wanted to compare correctness with a pure stock Hadoop run, we generated our input data with Hadoop (using the same Partitioner logic as the benchmark). We used the same data for both Hadoop runs of the microbenchmark and our M3R runs. This presents a challenge since although the pairs are subject to the same partitioner in M3R and Hadoop, the assignment of partitions to hosts is very different. M3R assigns partitions to hosts in a fixed manner, whereas Hadoop uses a dynamic approach. The host on which a given partition's data is stored is thus arbitrary, because it was written by the generator's reducer, which ran in Hadoop.

In M3R, M3R runs mappers on every host and a mapper is assigned, in the typical case, a local input split. This may not be the correct input split according to the partition/host mapping implied by partition stability, and thus pairs that are emitted with keys unmodified may end up being shuffled remotely. To avoid this, a 'repartitioner' job is run ahead of time, in M3R, using the identity mapper and reducer. This redistributes the HDFS storage of the data, using the shuffle, according to the M3R assignment of partitions to hosts. For the data described, this takes 83 seconds. This is a one-off cost, as the reorganized data can be used for any job, in any run of the benchmark subsequent to this.

To avoid this extra step when bringing Hadoop-output data into M3R, there are some ideas that we will pursue as further work. In the common case where the input data is partitioned along the same lines, but merely permuted across the hosts, HDFS remote reads could be used to bring the data into the correct mapper. The data would be cached in the right place so the cost would be only for the first iteration. This would be implemented using the PlacedSplit API, introduced in Section 4.3, to override M3R's preference of local splits. Additionally, if the data is evenly distributed, it may be possible to take the permutation forced on M3R by Hadoop's assignment of partitions to hosts, and re-use it, keeping those partitions stable throughout the M3R execution. This would avoid all network overhead. In general, however, the data might be partitioned completely differently, e.g. if the Hadoop job that produced the data had a different number of reducers than the M3R job. In these cases, a full repartition job is required, to re-arrange the data on a pair-by-pair basis.

## 6.2 Sparse Matrix Dense Vector Multiply

We discuss in detail the sparse vector multiply algorithm referred to in Section 3.

The sparse matrix $G$ is blocked into 1000 x 1000 blocks, using a custom key class that encapsulates a pair of ints as a two-dimensional index into the matrix. The value of such pairs is a compressed sparse column (CSC) representation of the sparse block. The dense vector $V$ is blocked into 1000 x 1 blocks. The same key type is used (with a redundant column value of 0) and each value is an array of double. When generating data, a sparsity value of 0.001 is used for the sparse matrix. The generation is done by a Hadoop program, and a repartitioning job is used to reorganize the data as described in Section 6.1. The pairs are partitioned using the row index. This means that a given partition will contain a number of rows of $G$ and matching blocks of $V$.

The algorithm consists of three iterations of calculating a new $V$ from $G$ and the previous $V$. Each iteration requires two jobs. The first job has a mapper for each of its $G$ and $V$ inputs. The $G$ mapper simply passes through each $G$ block, whereas the $V$ mapper broadcasts each $V$ block to every index of $G$ that needs to be multiplied by it (i.e. a whole column). The reducer receives each block of $G$ and associated $V$ and multiplies them. This yields a partial result of the corresponding new $V$ block, keyed by the index of the $G$ block that was used. To sum the results of a row of $G$, the second job collects them by using its map logic to rewrite the keys to have column 0. A single reduce call therefore receives all partial sums and can compute the new $V$ block, keyed by the row number and 0.

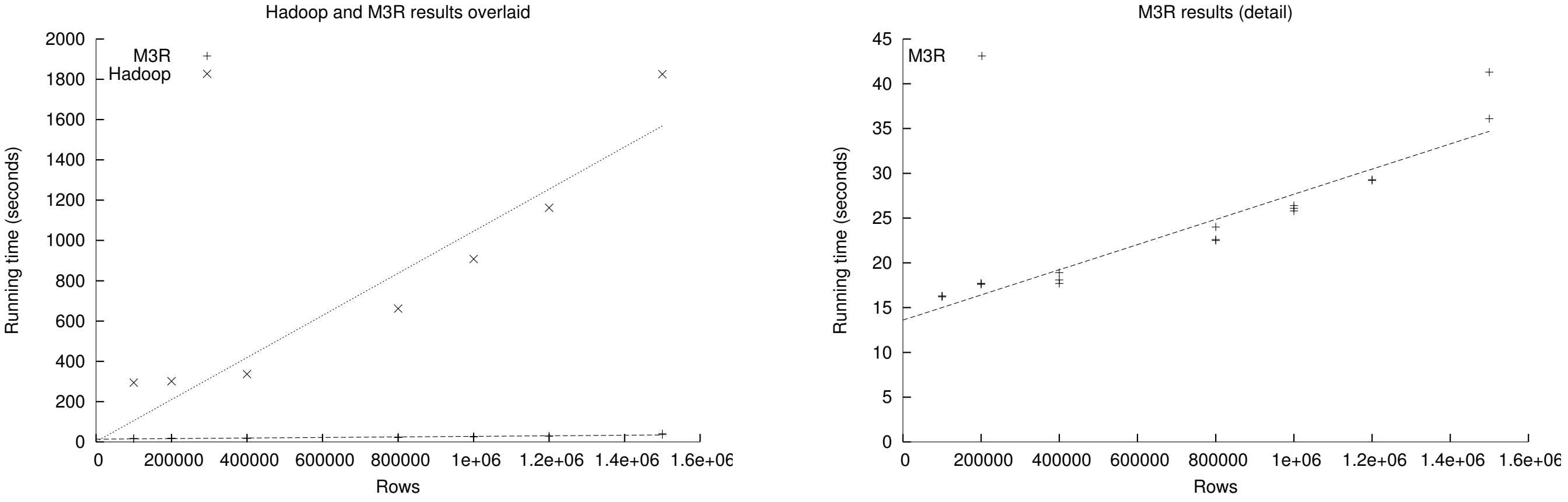

**Figure 7: Sparse matrix dense vector multiply**

All mappers and reducers are marked as producing only `ImmutableOutput`, allowing the M3R engine to use aliases wherever possible. Aside from the initial load, the only disk or network I/O performed is during the shuffle phase of the first job of each iteration, where the $V$ blocks are broadcast so that every host has a complete $V$ with which to multiply against its row of $G$. All communication that was not inherent to the multiplication operation, given the partitioning scheme we chose, has been eliminated. In order to make the application more representative of a real machine-learning algorithm, which would use many more than three iterations, we pre-populated our cache with the input data. This means that the initial I/O overhead (which if there were more iterations would be amortized across them) is not measured.

Figure 7 shows the comparison between M3R and Hadoop. The right hand graph shows just the M3R data, so its scalability is visible.

## 6.3 Word Count

Word count (Map Reduce's "Hello World") is an interesting case since none of M3R's optimizations apply. It is not an iterative job, so the cache does not come into play. It does not make use of partition stability. The vast majority of its shuffled pairs are remote. We modified the standard code to not mutate its pairs, and added the `ImmutableOutput` annotation to mapper and reducer. This means our instance of the performance profile is on the 100% end of the *Iteration* 1 line in Figure 6. However we cannot expect such great performance improvement over the HMR engine since that microbenchmark did not do any work on the keys, it only measured communication costs.

Figure 8 shows that the M3R engine is approximately twice as fast as HMR engine for these input sizes. Greater input sizes would still fit in the memory of these machines, but the M3R shuffle implementation currently has considerable memory overhead when large numbers of small pairs are used. De-duplication currently requires needs to compare each outgoing key/value pair against the previous ones, requiring the old ones to be kept around for longer then otherwise needed. We are planning to address this problem by relaxing X10 de-duplication to only check consecutive key/value pairs from the same mapper. This still allows the broadcast idiom of emitting pairs in a loop and does not require such a heavy-weight implementation.

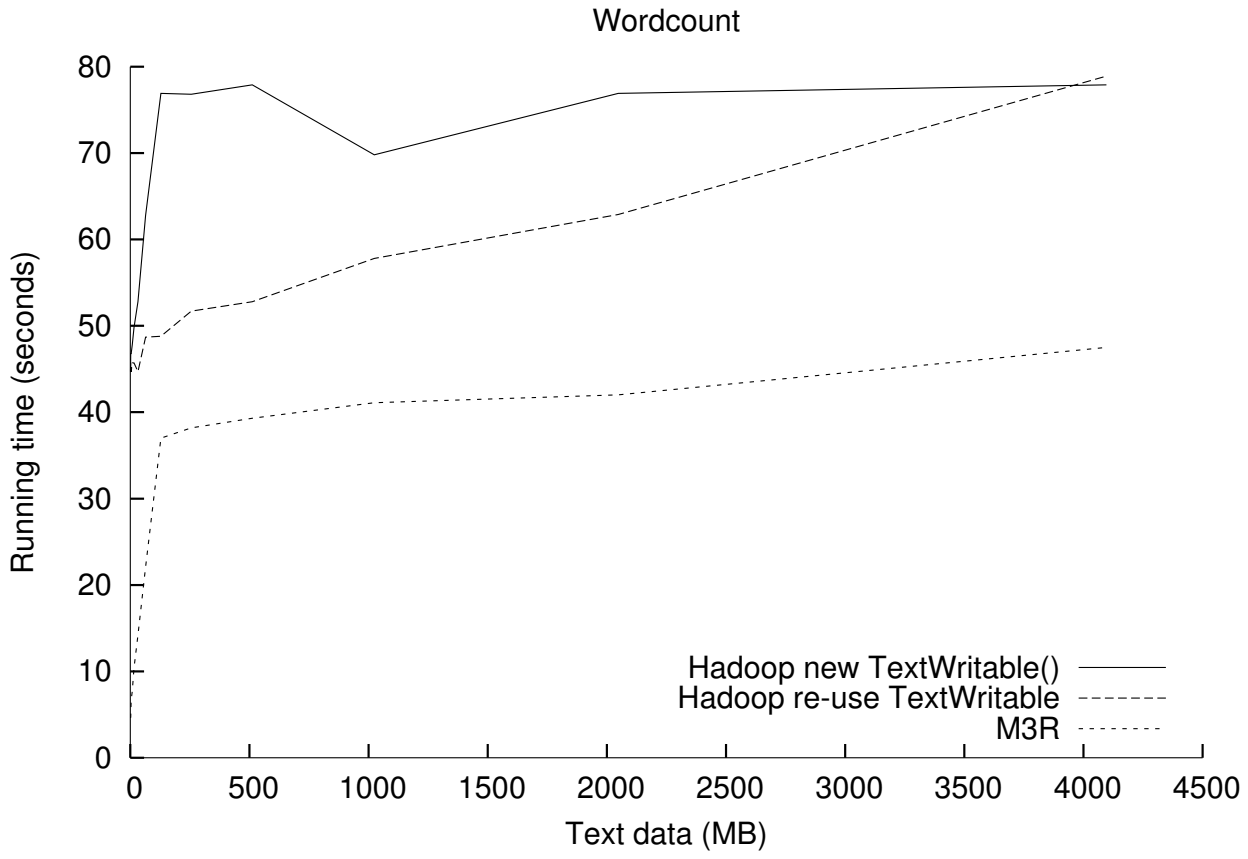

**Figure 8: Wordcount performance**

Also shown in Figure 8 is the performance cost of the modification to `WordCount` to allow use of `ImmutableOutput`. The version that allocates new `TextWritable` objects, and therefore can be annotated with `ImmutableOutput`, is subject to more memory pressure and GC churn. It is slower for small input sizes, but the gap disappears as the size increases.

## 6.4 System ML

System ML [15] is an R-like declarative domain specific language that permits matrix-heavy algorithms for machine learning to be written concisely and elegantly. The System ML compiler produces optimized Hadoop jobs. System ML is of interest in M3R benchmarking since it allows us to compare the performance of compiler-generated Map/Reduce code on M3R against Hadoop. Indeed, more generally Sys-tem ML offers a simple and convenient way to benchmark the performance of multiple Map Reduce implementations on standard Machine Learning algorithms.

Two minor changes (involving changes to a few lines of code) were made to the System ML compiler and runtime. (1) System ML modified some Hadoop classes to fix bugs. We had to port modifications to these classes we had made to the SystemML version. (2) The System ML runtime directly accessed some files in HDFS; these had to be modified to be M3R cache-aware.[3]

No modifications were made to the System ML compiler optimization algorithms. In particular, the code generated by the compiler is not aware of `ImmutableOutput` (hence is not optimized for cloning), and does not take advantage of partition-stability. Finally, the in-memory representation for sparse matrix blocks in the System ML runtime is about 10x less space-efficient than in the sparse matrix multiply code we wrote manually. These factors are not important for SystemML code run on Hadoop, but make a big difference in M3R. Thus, with appropriate modifications to the System ML compiler, we believe that much better numbers can be obtained on M3R, without compromising the numbers obtained on Hadoop.

We present performance results for three iterative matrix-based System ML programs. The matrices had a sparsity factor of 0.001 and were distributed with a blocking factor of 1000. (Note that System ML is capable of handling matrices with much larger sizes than the ones presented here.)

Performance results for **Global non-negative matrix factorization** are shown in Figure 9. The experiment varied the number of rows in `V`, keeping the number of columns constant at `100000`, and the width of `W` (height of `H`) was `10`. **Linear regression** performance is shown in Figure 10. The experiment varied the number of sample points, whereas the number of variables was constant at `10000`. **Page rank** performance is shown in Figure 11. The independent variable in this case was the size of the graph, i.e. the size of the square matrix `G`.

## 7. CONCLUSIONS

We have presented a new engine for Hadoop Map Reduce jobs, the M3R engine. This is aimed at a different design point than Hadoop – a design point that emphasizes in-memory, non-resilient execution and is therefore able to deliver substantially better performance than Hadoop on jobs that can fit in cluster memory.

In future work we plan to develop libraries of Map Reduce code, e.g. libraries for sparse matrix vector computations, that can run on the HMR engine (scaling to the size of cluster disks), while delivering very good performance for jobs that can fit in the size of cluster memory.

We also plan to develop X10-based M3R style engines (not necessarily based on Map Reduce) to provide fast in-memory performance for other APIs, such as APIs for sparse graphs, matrices, tables (cf HBase [18]) etc.

Finally, we believe it is possible to extend the M3R engine so that it can support resilience and elasticity. To support resilience, M3R wil need to detect node failure and recover by performing work proportional to the work assigned to the failed node. We believe this can be done in a more flexible way than that supported by HMR (which effectively checkpoints state to disk after every job). Similarly we believe it is possible to extend M3R to support elasticity – the ability to cope with a reduction or an increase in the number of places over which it is executing – without paying for it at the granularity of a single job (as HMR does).

[3]Since the file API is based on byte buffers, and the cache stores key-value pairs, these calls could not be trapped automatically. However, the System ML runtime immediately deserializes the data into key value pairs, hence we patched System ML to retrieve the pairs from the cache directly.

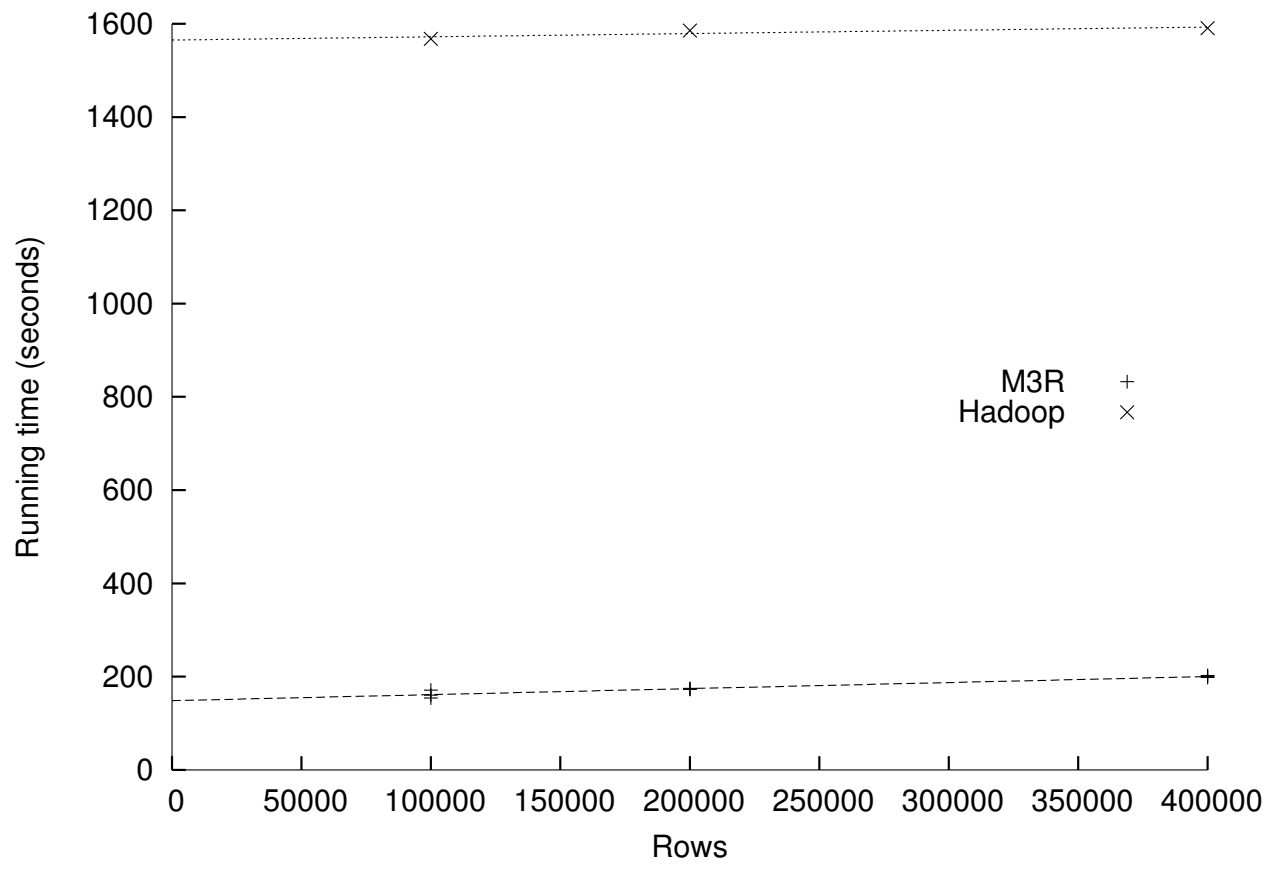

**Figure 9: System ML on Hadoop vs M3R: Global non-negative matrix factorization**

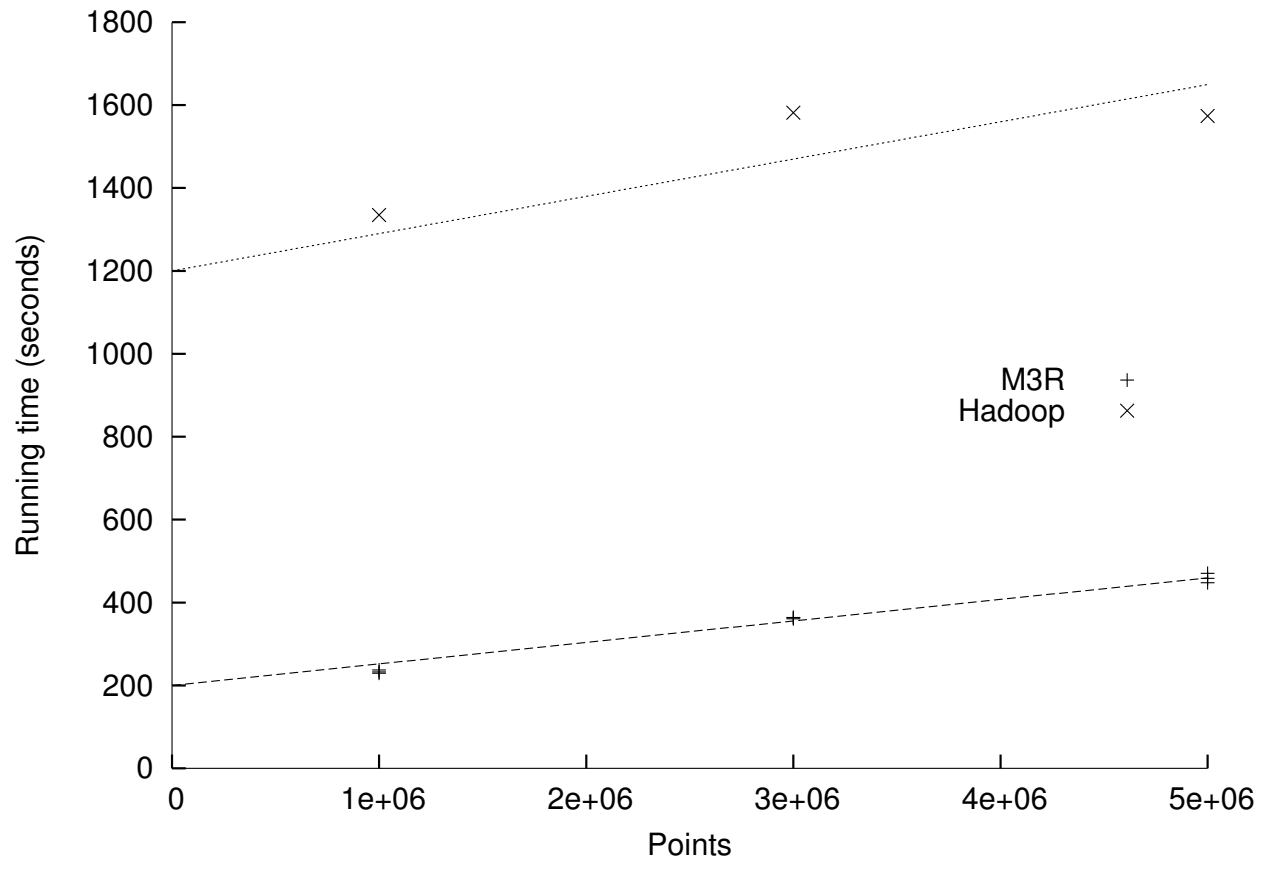

**Figure 10: System ML on Hadoop vs M3R: Linear regression**

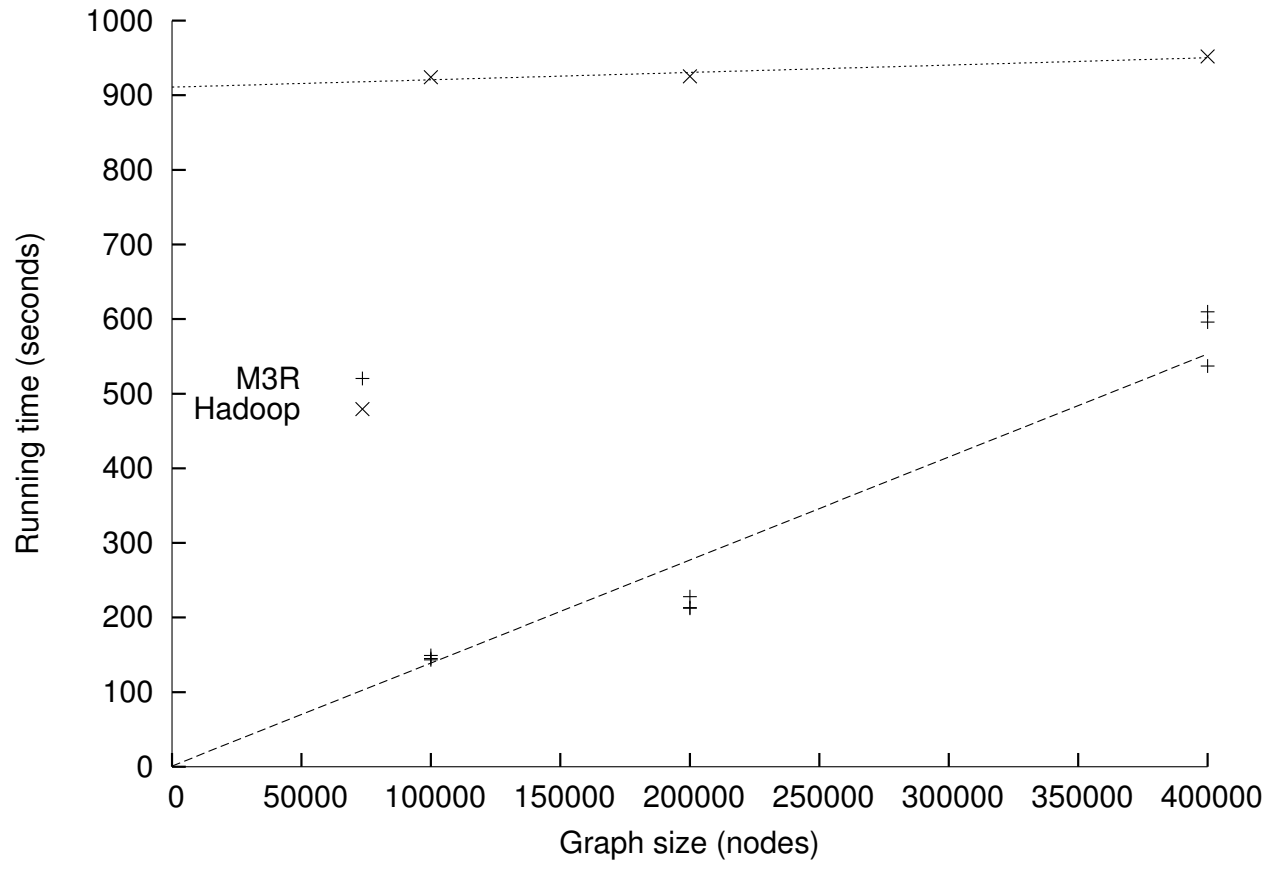

**Figure 11: System ML on Hadoop vs M3R: Page rank**

# 8. ACKNOWLEDGMENTS

Shiv Vaithyanathan helped create the concept of M3R. Shiv was particularly keen on compatibility with HMR APIs. His insistence has been a key foundation for the project. We thank Berthold Reinwald, Yuanyuan Tian, Doug Burdick, and Shirish Tatikonda for extensive discussions about System ML and for help in getting System ML to run on M3R. We also acknowledge valuable discussions with Amol Ghoting, Anju Kambadur and Vikas Sindhwani.

M3R has been developed in close collaboration with other members of the X10 team. We especially thank Yan Li, Dave Grove, Keith Chapman, Olivier Tardieu, Mikio Takeuchi, Salikh Zakarov and Hai Bo Lin. Yan implemented sparse matrix multiply on the core M3R engine. Dave and Keith were responsible for ensuring that the X10 serialization protocol worked with Hadoop. Mikio Takeuchi is primarily responsible for the implementation of the X10 Java interop that was crucial for this project. Mikio and Salikh are responsible for getting Jaql to run on M3R. Mandana Vaziri provided support with the Eclipse based development environment for X10, X10DT. We thank Steve Fink for his early contributions to the project.